\documentclass[aps,prd,preprint,nofootinbib]{revtex4}
\usepackage{epsf,epsfig,graphics,graphicx}
\usepackage{verbatim,color,ulem}
\newcommand{\be}{\begin{equation}}
\newcommand{\ee}{\end{equation}}
\newcommand{\bea}{\begin{eqnarray}}
\newcommand{\eea}{\end{eqnarray}}
\newcommand{\ba}{\begin{array}}
\newcommand{\ea}{\end{array}}
\usepackage{amssymb,amsmath,amsthm,graphicx}
\usepackage[mathscr]{eucal}
\usepackage{enumerate,color,verbatim,multirow,comment}
\begin{document}

\title{Out-of-Time-Order-Correlators in Holographic EPR pairs}
\author{Shoichi Kawamoto}
\email{ kawamoto_s@obirin.ac.jp} \affiliation{College of Arts and Sciences,
J.~F.~Oberlin University, Tokyo, Japan}
\author{Da-Shin Lee}
\email{dslee@gms.ndhu.edu.tw}
\author{Chen-Pin Yeh}
\email{chenpinyeh@gms.ndhu.edu.tw}
\affiliation{Department of Physics, National Dong Hwa University, Hualien, Taiwan, R.O.C.}

\begin{abstract}
In this note, we investigate the out-of-time-order correlators (OTOCs) for quantum fields in a holographic framework describing Einstein-Podolsky-Rosen (EPR) pairs. We compute the four-point and six-point OTOCs using the gravity dual, represented by the string worldsheet theory in Anti-de Sitter (AdS) space. These correlators quantify the rate at which information is scrambled, leading to the disentanglement of the EPR pair.
We demonstrate consistency between two approaches for calculating OTOCs: the holographic influence functional on worldsheets perturbed by shock waves, and the worldsheet scattering in the eikonal approximation. We show that the OTOCs exhibit an initial phase of exponential growth, with six-point correlators indicating a marginally longer scrambling time compared to four-point correlators.
\end{abstract}

\maketitle

\section{introduction}
There is evidence suggesting that in theories of quantum gravity, quantum information can spread at a maximal speed \cite{Sekino_08, Maldacena_16}. This gives a hint of how a theory of quantum gravity should be constructed. Notably, in the holographic framework of gravity (AdS/CFT), this maximal speed has been explicitly demonstrated \cite{Shenker_15}. A tool to probe this rapid spread of quantum information (namely, the butterfly effect in quantum chaos) is the out-of-time-order-correlators (OTOCs). For example, the initial exponential decay of four-point OTOCs $\langle\hat A(t)\hat B(0)\hat A(t)\hat B(0)\rangle$, observed during times well before the scrambling time but after thermalization, implies very sensitive dependence of the measurement $\hat A$ at time $t$ to the perturbation $\hat B$ at time zero. This behavior is an indication of quantum chaotic behavior \cite{Roberts_14}.

In the AdS/CFT correspondence, there is a close connection between geometry and quantum information. One example is the ER=EPR conjecture \cite{ER=EPR}, which demonstrates the relation between wormhole geometry (ER bridge) and the quantum entangled (EPR) pair. In previous work \cite{Yeh_23}, we examined this conjecture using a holographic model of an EPR pair, represented by a string in AdS space with endpoints anchored on the AdS boundary, accelerating in opposite directions. In this scenario, the induced metric on the string's worldsheet forms an AdS wormhole. By perturbing this wormhole with shock waves and analyzing the field correlators between the EPR pair, we observed potential effects on their quantum entanglement. In the present work, we aim to further characterize this influence by examining the fields' OTOCs.

It turns out that in the AdS/CFT correspondence, the four-point OTOCs can be obtained from either the AdS bulk eikonal scattering amplitudes \cite{Shenker_15} or the two-point correlators in AdS with shock wave backgrounds \cite{Shenker_13}. In flat spacetime, the equivalence between eikonal quantum gravity scattering amplitudes and two-point correlators in a shock wave background is well established \cite{Kabat_92}. A similar equivalence can also be shown in AdS gravity \cite{Cornalba_07}. In \cite{Boer_17}, the four-point OTOCs for a holographic EPR pair were obtained by relating them to the scattering of worldsheet perturbations. However, it remains unclear whether the same equivalence between four-point OTOCs and two-point correlators in a shock wave background holds in this case; this is because the string worldsheet theory is not truly gravitational, while the proofs of equivalence in \cite{Kabat_92} and \cite{Cornalba_07} rely on the gravitational actions. Following the methodology of \cite{Shenker_13}, we will argue in this paper that the equivalence still works due to the AdS/CFT dictionary we employ. 

Here, we summarize the main results of this paper and their relations to other works. In \cite{Yeh_23}, we calculated the two-point correlators between a holographic EPR pair from the worldsheet with one and two shock waves, respectively. In this paper, we reinterpret the results as four-point and six-point OTOCs (similar results were also obtained in \cite{Murata_17,Banergee_19} by using different methods). The new results here are the calculations of four-point and six-point OTOCs using the worldsheet scattering amplitude in the eikonal approximation (see equations (\ref{4OTOC}) and (\ref{F6})).  In \cite{Boer_17}, the four-point OTOC in the holographic EPR setup was also obtained by using the worldsheet scattering theory. However, no prior study has identified a direct connection between OTOC calculations derived from shock wave backgrounds and those from eikonal scattering amplitudes in this setting. We establish the equivalence by identifying the two-to-two eikonal scattering amplitude with the doubling scattering parameter (see equation (\ref{amplitude})). This identification clarifies the connection between wormhole geometry and quantum information in the spirit of the ER=EPR conjecture. We confirm this argument by comparing both the four-point and six-point OTOCs from eikonal scattering amplitudes on the worldsheet wormhole and those from two-point correlators in worldsheet shock wave backgrounds as calculated in \cite{Yeh_23}. In parameter regions where the results from both methods have analytical forms, we find complete agreement. In a different setting \cite{Haehl_21}, six-point OTOCs from AdS/CFT have been used to probe the black hole interior. It would be interesting to know if the worldsheet wormhole in the current work can also be probed in this way. In \cite{Giombi_23,Giombi_24}, they also considered the string worldsheet in AdS space and calculate the four-point and six-point OTOCs. However, they used different string profiles and interpreted the results as scalar correlators on the Wilson line defect in the boundary CFT (instead of the holographic EPR pair). Their results have been used to clarify the structure of the boundary CFT. It will also be interesting to know how it can be related to the six-point OTOCs found here.

In the next section, we review the setup for the holographic EPR pair and discuss the correlators derived from the holographic influence functional method in shock wave backgrounds \cite{Yeh_23}. We also point out the connection between these results and the OTOCs. Subsequently, in Section \ref{scattering}, we employ the worldsheet scattering amplitude in the eikonal approximation to compute the four-point and six-point OTOCs and compare these results with those obtained from two-point functions in shock wave backgrounds. Finally, we conclude with a discussion of potential future research directions.

\section{Worldsheet wormholes with shock waves}

In this section, we first review the holographic setting for EPR pairs. The bulk description involves a string moving in $AdS_{d+1}$ space, with its two ends anchored at the $AdS$ boundary. To examine the relation between entanglement and wormhole geometry (the ER=EPR conjecture), we consider a case where the two ends of the string accelerate uniformly in opposite directions, such that the induced metric on the string's worldsheet describes an AdS wormhole. We then analyze the backreaction on the EPR pair caused by shock waves on the string worldsheet and calculate the cross-correlator between the EPR pair. Finally, we identify the cross-correlators in the shock wave background with the out-of-time-ordered correlators (OTOCs) in thermalized states.

\subsection{Accelerating Strings}

\begin{figure}[h]
\centering
\includegraphics[scale=1]{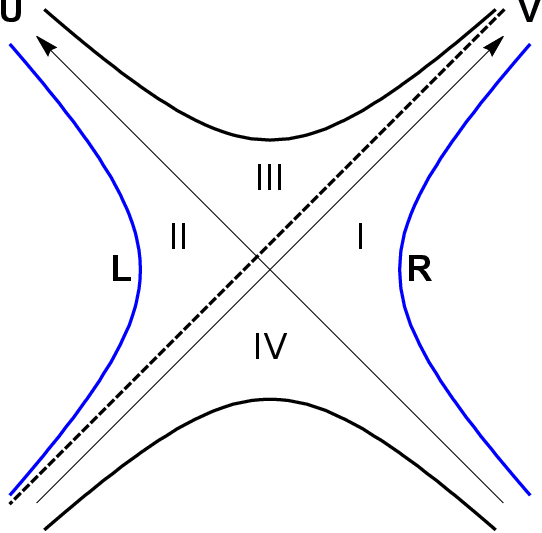}
\caption{The worldsheet metric in $(U,V)$-coordinate. Both $U$ and $V$ have their ranges from $-\infty$ to $\infty$. The left (right) boundary is identified with the left (right) quark trajectory, and are described by $UV=-1$. A shock wave trajectory is illustrated by the dashed straight line.}
\label{kruskal}
\end{figure}

We consider the string worldsheet theory as described in \cite{Yeh_23}. The background $AdS_{d+1}$ space is parameterized by the metric\footnote{Even though we consider general $d$ here, the results in the following discussion do not depend on $d$. The reason is that, in the linear response region of  the worldsheet theories we consider below, the fluctuations in different directions decouple and we concentrate only on the motion in one direction.},
  \be
  \label{ads}
  ds^2=\frac{R^2}{z^2}(-dt^2+dz^2+dx^2+\sum_{i=1}^{d-2}dy_i^2)\, ,
  \ee
where $R$ sets the curvature scale and the cutoff boundary is located at $z=z_m$. The surface $z=\infty$ represents the Poincar\'e horizon, which is the boundary of Poincar\'e patch within the global AdS space. According to the dictionary of the AdS/CFT correspondence, quantum gravity in this background (bulk) is dual to the conformal field theory (CFT) living at $z=z_m$ (boundary), which is a Minkowski spacetime parameterized by $(t,x,y_1,y_2,...,y_{d-2})$. A classical string moving in this $AdS_{d+1}$ background has an exact solution where its two ends reside on the boundary, accelerating uniformly in opposite directions along $x$-axis with acceleration $b^{-1}$. In this case, the induced metric on the string's worldsheet is given by \cite{Yeh_23}:
    \be
    \label{ws}
    ds_{ws}^2=\frac{1}{(t^2+b^2-x^2)^2}\left((x^2-b^2)dt^2-2txdtdx+(t^2+b^2)dx^2\right)\, .
    \ee
This metric also describes a two-sided AdS black hole (thus a wormhole) with inverse temperature $\beta=2\pi b$ and bifurcate horizons at $x=\pm t$. The two boundaries of this AdS black hole, given by $x=+\sqrt{t^2+b^2-z_m^2}$ (right boundary) and $x=-\sqrt{t^2+b^2-z_m^2}$ (left boundary), correspond to the trajectories of entangled pair in the boundary theory. In the left outside region of the worldsheet wormhole (region II in figure \ref{kruskal}), we define the light-like coordinates,
 \be
 u=\frac{-b^2t+xb\sqrt{b^2+t^2-x^2}}{x^2-t^2},~~~v=\frac{-b^2t-xb\sqrt{b^2+t^2-x^2}}{x^2-t^2} \, ,
 \ee
with $-\infty<v<u<\infty$. It is convenient to represent the wormhole in the Kruskal coordinates defined as\footnote{%
$U$ and $V$ here are related to those in \cite{Yeh_23}
as $U_\text{here}=\tan \frac{V_\text{there}}{2b}$ and $V_\text{here}=\tan \frac{U_\text{there}}{2b}$.},
  \be
  U=e^{-\sinh^{-1}\frac{u}{b}},~~~V=-e^{\sinh^{-1}\frac{v}{b}} \, ,
  \ee
we then have the following wormhole metric
  \be
  \label{ws2}
  ds_{ws}^2=\frac{-4dUdV}{(1+UV)^2} \, .
  \ee
The same metric can be obtained in all other regions of the wormhole through similar coordinate transformations. The bifurcate horizons are at $U=0$ and $V=0$, while the trajectories of EPR pair correspond to $UV=-1$. The Poincar\'e horizons are at $UV=1$ (see figure \ref{kruskal}).

\subsection{Shock Wave Backgrounds}
In \cite{Yeh_23}, the worldsheet wormhole is distorted by massless energy quanta sent from the left boundary, propagating along $U\simeq0$ (see figure \ref{kruskal}). In the ``double scaling limit'', the shock wave geometry is obtained by patching two worldsheets along $U\simeq0$ with a shift in the $V$-coordinate. Specifically, after crossing the shock wave from the left to the right region, the coordinates transform from $(U,V)$ to $(U',V')$,
where $U=U'$ and
 \be
 \label{shift}
V=V'+ \gamma \, ,
 \ee
with $\gamma$ being a parameter characterizing the strength of the shock wave. The double scaling limit is obtained by considering two worldsheets like (\ref{ws}) with different acceleration (say $b^{-1}$ and $b'^{-1}$) that are patched together along the trajectory of the shock wave emitted at $u=u_0$ at the boundary of region II. We keep
 \be
\gamma\equiv \frac1{2b}(b-b')e^{-\sinh^{-1}\frac{u_0}{b}}
\ee
fixed but let $b'\to b$ and $u_0\to -\infty$.  In this limit, the energy of the shock becomes small and the time when the shock was emitted becomes early. So from the boundary point of view, the large blue shift factor compensates the smallness of energy and we have a finite shock wave effect characterized by $\gamma$. Using the holographic influence functional method and in the small $\gamma$ limit\footnote{Note that the shock wave background is valid for arbitrary $\gamma$. However we were only able to obtain an analytic form of the influence functional in leading order of $\gamma$.}, \cite{Yeh_23} derived the cross-correlator between fields couple to left and right EPR particles respectively;
 \bea
 \label{grl}
 &&\langle \gamma|\hat F_L(t)\hat F_R(t')|\gamma\rangle \equiv G_{LR}(t,t')\simeq\frac{b}{\sqrt{b^2+t^2-z_m^2}}\frac{b}{\sqrt{b^2+(t'+b\gamma)^2-z_m^2}}\nonumber\\
 &&\times\int_{-\infty}^{\infty}\frac{d\omega}{2\pi}G_{LR}(\omega)e^{-i\omega (\tau(t)+ \tau(t'+b\gamma))} \, ,
 \eea
where $\tau(t)=b\tanh^{-1}\frac{t}{\sqrt{b^2+t^2-z_m^2}}$, $|\gamma\rangle$ is the boundary state dual to the shock wave geometry, and
 \be
G_{LR}(\omega)=\frac{2}{e^{\frac{\omega}{2T}}(1-\frac{i\gamma\omega}{2\pi T})-e^{-\frac{\omega}{2T}}}i\mbox{Im} G_r(\omega) \, ,
 \ee
 where
 \be
G_r(\omega)=\frac{R^2T_0}{2b^2}\left(\frac{b^2-z_m^2}{z_m}\frac{\omega^2}{1+\omega^2z_m^2}+i\frac{\omega+b^2\omega^3}{1+\omega^2z_m^2}\right) \, ,
\ee
is identified with the retarded Green function for the field $\hat F$. Here, $T_0$ is the string tension. In the AdS/CFT dictionary, the operator $\hat F$ is sourced by the quantum field associated with string fluctuations.  However, it should also be noted that $\langle \gamma=0|\hat F_L(t)\hat F_L(t')|\gamma=0\rangle$ takes the form of thermal time-order correlator in finite temperature $\beta$ with $G_{LL}(\omega)=\mbox{Re}G_r(\omega)+i\mbox{Im}G_r(\omega)\coth\frac{\beta\omega}2$ \cite{Yeh_23}. Thus, we can identify $|\gamma=0\rangle$ with the thermal field double state $|TFD\rangle$, which is consistent with the conjecture in \cite{Maldacena_01}.

\subsection{Relation to OTOCs}

In the context of AdS/CFT, the relationship between free propagators in shock wave backgrounds and thermal OTOCs was discussed in \cite{Shenker_15}. The same relationship also applies in our setting due to the identification, $|\gamma=0\rangle=|TFD\rangle$ as discussed in the previous subsection.
Let us consider the shock wave state created by the operator $\hat W_L(t_0)$, so that $|\gamma\rangle=\hat W_L(t_0)|TFD\rangle$. Here, the shock wave is emitted from the left particle at $t_0$ (in the double scaling limit, $t_0\rightarrow -\infty$).
The correlator $G_{LR}(t,t')$ is then given by
\be
\label{glr}
G_{LR}(t,t')=\langle TFD|\hat W^{\dagger}_L(t_0)\hat F_L(t)\hat F_R(t')\hat W_L(t_0)|TFD\rangle
\ee
This expression can be related to the thermal OTOCs by using the definition, $|TFD\rangle=Z_{\beta}^{-1/2}\sum_n e^{-\frac{\beta}{2}E_n}|n\rangle_L\otimes|n\rangle_R$, where $Z_{\beta}=\sum_n e^{-\beta E_n}$ is the partition function. We also define $\hat O_L=\hat O\otimes\hat 1$ and $\hat O_R=\hat 1\otimes \hat O^{T}$ (It is $\hat O^T$ on the right boundary, since the bulk time-like isometry acts in opposite $t$-direction on two boundaries. So the total Hamiltonian is $\hat H\otimes \hat 1-\hat 1\otimes \hat H$). We then have
  \bea
  \label{glra}
  &&G_{LR}(t,t')=Z_{\beta}^{-1}\sum_{mn}\langle n|\hat W^{\dagger}(t_0)\hat F(t)\hat W(t_0)|m\rangle\langle n|e^{-\frac{\beta}{2}\hat H}\hat F^T(t')e^{-\frac{\beta}{2}\hat H}|m\rangle\nonumber\\
  &&=Z_{\beta}^{-1}\sum_{mn}e^{-\beta E_n}\langle n|\hat W^{\dagger}(t_0)\hat F(t)\hat W(t_0)|m\rangle\langle m|\hat F(t'-\frac{\beta}{2}i)|n\rangle\nonumber\\
  &&=Z_{\beta}^{-1}\sum_{n}e^{-\beta E_n}\langle n|\hat W^{\dagger}(t_0)\hat F(t)\hat W(t_0)\hat F(t'-\frac{\beta}{2}i)|n\rangle\nonumber\\
  &&\equiv \langle\hat W^{\dagger}(t_0)\hat F(t)\hat W(t_0)\hat F(t'-\frac{\beta}{2}i)\rangle_{\beta}
  \eea
which is the thermal OTOC in complexified time. By a similar derivation, other types of four-point OTOCs can also be related to $G_{LR}(t,t')$ through some analytical continuations.

\section{Worldsheet scattering in eikonal approximation}
\label{scattering}

Since the result for OTOCs we obtained via the shock wave method in \cite{Yeh_23} is only valid in the small-$\gamma$ limit, we seek an alternative approach for calculating OTOCs. Here, we demonstrate the equivalence between the shock wave method and the eikonal scattering amplitude approach. This equivalence directly shows that the shock wave background encodes how quantum information is scrambled in the boundary theory.

We begin by using the eikonal approximation to compute the four-point OTOCs and compare the results with those obtained from free propagators in a single-shock-wave background. Finally, at the end of this section, we calculate the six-point OTOCs and perform a similar comparison with free propagators in a two-shock-wave background.

 \subsection{The Setting}

In the AdS/CFT correspondence, the operator $\hat F(t)$ is dual to the worldsheet perturbation $q(U,V)$, which can be obtained from the boundary data $q_0(t)=q(z(U,V)=z_m)$, where $z(U,V)=b\frac{1+UV}{1-UV}$. In the Lorentzian formalism, as in \cite{Yeh_23}, we also need the boundary conditions at the horizons. The relation is given by
   \be
   q(U,V)=\int dt \langle q(U,V)\hat F(t)\rangle q_0(t) \,,
   \ee
where $\langle q(U,V)\hat F(t)\rangle$ is the bulk-to-boundary correlator that satisfies the linearized equation of motion for worldsheet perturbations with a delta function source at the boundary, and $q_0(t)$ acts as the source for $\hat F$. According to the AdS/CFT dictionary, in large $N$ and small perturbation limit,
    \be
    \hat F(t)|TFD\rangle=\int dV \langle q(U=0,V)\hat F(t)\rangle |V\rangle
    \ee
where $|V\rangle$ is the position eigenstate for worldsheet perturbations at $U=0$ slice. Note that the dictionary can be formulated on any bulk slice, and $U=0$ is a convenient choice for matching shock wave calculations. To calculate the high energy forward (eikonal) scattering of worldsheet fields, it is also convenient to formulate the bulk-to-boundary correlators in momentum space. Instead of directly calculating the correlator in (\ref{glr}), we calculate the correlator (following \cite{Shenker_15}),
 \be
 \langle TFD|\hat W^{\dagger}_L(t_1)\hat F_L(t_2)\hat W_L(t_3)\hat F_L(t_4)|TFD\rangle=\langle\hat W^{\dagger}(t_1)\hat F(t_2)\hat W(t_3)\hat F(t_4)\rangle_{\beta} \,,
 \ee
which is related to $G_{LR}(t,t')$ through analytical continuation from (\ref{glra}). We define the related momentum space wavefunctions as
  \bea
  &&\phi_2(p_2^U)=\int dV e^{2ip_2^UV}\langle q(U=0,V)\hat F_L^{\dagger}(t_2)\rangle \,,\\
  &&\phi_4(p_4^U)=\int dV e^{2ip_4^UV}\langle q(U=0,V)\hat F_L(t_4)\rangle \,.
  \eea
Similarly, for the operator $\hat W$, we assume that there is the dual worldsheet field $q_W(U,V)$, and define
    \bea
  &&\phi_1(p_1^V)=\int dU e^{2ip_1^VU}\langle q_W(U,V=0)\hat W_L(t_1)\rangle \,, \\
  &&\phi_3(p_3^V)=\int dU e^{2ip_3^VU}\langle q_W(U,V=0)\hat W_L(t_3)\rangle \,.
  \eea
Using the AdS/CFT dictionary, we then represent the initial and final states as
  \bea
  &&|\psi_{in}\rangle\equiv\hat W_L(t_3)\hat F_L(t_4)|TFD\rangle=\int\int dp_3^Vdp_4^U\phi_3(p_3^V)\phi_4(p_4^U)|p_3^Vp_4^U\rangle_{in} \,,\\
  &&|\psi_{out}\rangle\equiv\hat F^{\dagger}_L(t_2)\hat W_L(t_1)|TFD\rangle=\int\int dp_1^Vdp_2^U\phi_1(p_1^V)\phi_2(p_2^U)|p_1^Vp_2^U\rangle_{out} \,,
  \eea
where $|p^Vp^U\rangle$ is the tensor product of the momentum eigenstates of $q$ and $q_W$ quanta, with the normalization $\langle p|p'\rangle=\frac{4}{\pi}p\delta(p-p')$.
The eikonal approximation boils down to assuming that $|p_1^Vp_2^U\rangle_{out}\simeq e^{i\delta(s)}|p_3^Vp_4^U\rangle_{in}$, where $\delta(s)$ is the two-to-two scattering amplitude, obtained by summing only ladder diagrams, and is a function of the center-of-mass energy $s=-g_{UV}p_1^Vp_2^U$. Here, $g_{UV}=-4$ is the horizon value of the worldsheet metric in (\ref{ws2}).
Now we can represent the cross correlator, which is related to the thermal OTOC, as the overlap of bulk wavefunctions
  \bea
  \label{overlap}
  &&\langle TFD|\hat W^{\dagger}_L(t_1)\hat F_L(t_2)\hat W_L(t_3)\hat F_L(t_4)|TFD\rangle=\langle \psi_{out}|\psi_{in}\rangle\nonumber\\
  &&=\frac{16}{\pi^2}\int\int dp^Vdp^Ue^{i\delta(s)}p^V\phi_1^*(p^V)\phi_3(p^V)p^U\phi_2^*(p^U)\phi_4(p^U) \,.
  \eea

\subsection{Relation to the Two-point Functions in Shock Wave Backgrounds}

Before evaluating the integration (\ref{overlap}) for the worldsheet model, we first see its relation to the calculation of correlators in shock wave backgrounds. The cross-correlator can be represented as a bulk integral over the on-shell worldsheet perturbations $q_2(U,V)$ and $q_4(U,V)$, which are dual to $\hat F_L(t_2)$ and $\hat F_L(t_4)$, respectively. For a single-shock-wave background in double scaling limit, where the shift in the $V$ coordinate at the $U=0$ surface is described in (\ref{shift}), the on-shell action becomes (see also \cite{Yeh_23})
 \bea
 &&\langle \gamma|\hat F_L(t_2)\hat F_L(t_4)|\gamma\rangle=2i\int dV q_2^*(U=\epsilon,V)\partial_V q_4(U=-\epsilon,V)\nonumber\\
 &&=2i\int dV  \left(\int dp_2^U\phi_2^*(p_2^U)e^{2ip_2^UV}\right)\partial_V \left(\int dp_4^U\phi_4(p_4^U)e^{-2ip_4^U(V-\gamma)}\right)\nonumber\\
 &&=2i\int dp^U e^{2ip^U\gamma}p^U\phi_2^*(p^U)\phi_4(p^U) \,.
 \eea
Compare this expression with the one in (\ref{overlap}), we find that if we assume $q_W$ field is much heavier than $q$ field (or equivalently the $\hat W$ operator have a much larger dimension than the $\hat F$ operator), then the $p^V$ integration in (\ref{overlap}) can be performed by the saddle-point approximation, and the remaining $p^U$ integration is exactly the one in the shock wave calculation when we identify
 \be
 \label{amplitude}
 \delta(s)=2p^U\gamma \,.
 \ee
To derive (\ref{amplitude}), we need a microscopic model describing the interaction of $q_W$ quanta to the string worldsheet perturbations. Here, we will take a bottom-up approach by just assuming that, to leading order in linearized theory, $\gamma\propto p^V$.
In \cite{Yeh_23}, the double scaling parameter, $\gamma$ was estimated to be proportional to the total energy of the shock wave, consistent with the assumption here.
Since we also have $s=4p^Up^V$, the assumptions imply that $\delta(s)=\gamma's$, where $\gamma'\equiv \gamma/2p^V$ is a constant independent of $p^U$ and $p^V$. In summary, we have the representation of the cross correlator as
  \be
  \label{overlap2}
  \langle \psi_{out}|\psi_{in}\rangle=\frac{16}{\pi^2}\int\int dp^Vdp^Ue^{i\gamma's}p^V\phi_1^*(p^V)\phi_3(p^V)p^U\phi_2^*(p^U)\phi_4(p^U) \,.
  \ee

\subsection{Four-point OTOCs}

In the linearized worldsheet theory, due to the isometry of (\ref{ws2}), the bulk-to-boundary correlator for any operator $\hat O$ takes the form (after subtracting the divergent term $\propto \ln z_m$)
 \be
 \langle \phi_{Q_L}(U,V)\hat O_L(t)\rangle=C_O e^{-\Delta_O D}=C_O\left(\frac{1+UV}{2(1+Ve^{\sinh^{-1}t/b}-Ue^{-\sinh^{-1}t/b}-UV)}\right)^{\Delta_O} \,,
 \ee
 where $D$ is the geodesic distance between the point $(U,V)$ and the boundary point at time $t$, and $C_O$ and $\Delta_O$ are constants depending on the form of $\hat O$.
We also have $\langle \phi_{Q_L}(U,V)\hat O_L(t)\rangle=\langle \phi_{Q_R}(U,V)\hat O_R(-t)\rangle$ and $\langle \phi_{Q_L^{\dagger}}(U,V)\hat O_L^{\dagger}(t)\rangle=\langle \phi_{Q_L}(U,V)\hat O_L(t)\rangle^*$. By applying the extrapolation dictionary of the AdS/CFT correspondence,
 \be
 \label{cross}
 \langle TFD|\hat O_L(t)\hat O_R(0)|TFD\rangle
\stackrel{t\gg b}{\longrightarrow} C_O\left(\frac{t}{b}\right)^{-\Delta_O} \, ,
 \ee
 to the operator $\hat F$ and comparing with $\langle TFD|\hat F_L(t)\hat F_R(0)|TFD\rangle$ we calculated in \cite{Kawamoto_22}, we find that
 \be
 \label{CF}
 C_F=\frac{16R^2T_0i}{b^4}, \qquad \Delta_F=2 \,,
 \ee
where $R$ is the ambient AdS curvature radius, $T_0$ is the string tension, and $b=\frac{\beta}{2\pi}$.
For convenience in the following calculations, we define $\tau_i=b\sinh^{-1}\frac{t_i}{b}$, which is the proper time for the accelerating particles in the boundary theory (and $t$ is the lab frame time for the EPR pair). Now, we can calculate the wavefunction,
  \be
  \label{phi2f}
  \phi_2(p^U)=\int dV e^{2ip^UV}C_F^*\left(\frac{1}{2+2Ve^{\tau^*_2/b}}\right)^{\Delta_F}
  \ee
by using the contour integration.
Assuming Im$\tau_2<0$, we then have
   \be
   \label{phi2}
   \phi_2(p^U)=\Theta(p^U)\frac{\pi i C_F^*e^{-\tau_2^*/b}}{\Gamma(\Delta_F)}\left(ip^Ue^{-\tau_2^*/b}\right)^{\Delta_F-1}e^{-2ip^Ue^{-\tau^*_2/b}} \,.
   \ee
Similarly,
  \bea
  &&\phi_4(p^U)=\Theta(p^U)\frac{ \pi i C_Fe^{-\tau_4/b}}{\Gamma(\Delta_F)}\left( ip^Ue^{-\tau_4/b}\right)^{\Delta_F-1}e^{-2ip^Ue^{-\tau_4/b}}\label{phi4} \,, \\
  &&\phi_1(p^V)=\Theta(p^V)\frac{-\pi i C_We^{\tau_1/b}}{\Gamma(\Delta_W)}\left(-ip^Ve^{\tau_1/b}\right)^{\Delta_W-1}e^{2ip^Ve^{\tau_1/b}}\label{phi1} \,, \\
  &&\phi_3(p^V)=\Theta(p^V)\frac{-\pi i C_We^{\tau_3/b}}{\Gamma(\Delta_W)}\left(-ip^Ve^{\tau_3/b}\right)^{\Delta_W-1}e^{2ip^Ve^{\tau_3/b}} \,,
  \eea
by assuming Im$\tau_4>0$, Im$\tau_1>0$, and Im$\tau_3>0$.
We use these wave functions and the identification that the eikonal amplitude is proportional to the center-of-mass energy, $\delta(s)=\gamma's$.
By defining $\tau_1=\tau_0+ib\epsilon_1$, $\tau_3=\tau_0+ib\epsilon_3$, $\tau_2=ib\epsilon_2$, $\tau_4=ib\epsilon_4$, and after redefining of the integral paths $p=p^V2i\epsilon_{13}e^{\tau_0/b}$, $q=p^U2i\epsilon_{24}$ with $\epsilon_{13}= e^{-i\epsilon_1}- e^{i\epsilon_3}$, $\epsilon_{24}=-e^{-i\epsilon_2}+e^{-i\epsilon_4}$, the integration in (\ref{overlap2}) is found to be
  \bea
  &&\langle\psi_{out}|\psi_{in}\rangle=\frac{16\pi^2|C_W|^2C_F^2}{\Gamma^2(\Delta_W)\Gamma^2(\Delta_F)}\left(\frac{1}{16\sin^2\frac{\epsilon_1-\epsilon_3}{2}}\right)^{\Delta_W}\left(\frac{1}{16\sin^2\frac{\epsilon_2+\epsilon_4}{2}}\right)^{\Delta_F}\\
  &&\times\int_0^{\infty}\int_0^{\infty}dpdq p^{2\Delta_W-1}q^{2\Delta_F-1}e^{-p-q}e^{-i\gamma'e^{-\tau_0/b}pq/\epsilon_{13}\epsilon_{24}} \,.
\label{overlap3}
  \eea
We now assume that the mass of $q_W(U,V)$ field is much larger than that of the worldsheet perturbation $q(U,V)$, or equivalently, $\Delta_W \gg \Delta_F=2$.
The $p$-integration can thus be evaluated by the saddle-point approximation, dominated by the region where $p\simeq 2\Delta_W$.
The remaining $q$-integration becomes
  \bea
  &&\langle \psi_{out}|\psi_{in}\rangle\propto\int_0^{\infty}dq q^{2\Delta_F-1}e^{-q}e^{-i\gamma'e^{-\tau_0/b}2\Delta_Wq/\epsilon_{13}\epsilon_{24}}\\
  &&\propto \left(1+i\gamma'e^{-\tau_0/b}\frac{2\Delta_W}{\epsilon_{13}\epsilon_{24}}\right)^{-2\Delta_F} \,.
  \eea
To compare this result with $G_{LR}(0,0)$ obtained using the shock wave method (which is identified as the thermal OTOC from equation (\ref{glra})), we set $\epsilon_2=0$, $\epsilon_4=\frac{\beta}{2b}$, leading to $\epsilon_{24}=-2$. We also take $\epsilon_1=\epsilon_3=\frac{z_m}{b}$ which is a natural cutoff in the system, and this gives $\epsilon_{13}=-\frac{2iz_m}{b}$. Hence, by using the eikonal scattering method, we obtain the thermal OTOC
  \be
  \label{4OTOC}
  \langle\hat W(t_0)\hat F(0)\hat W(t_0)\hat F(0)\rangle_{\beta}\propto \left(1+b\bar{\gamma}e^{-\tau_0/b}\right)^{-2\Delta_F}
  \ee
where $\bar{\gamma}\equiv\gamma'\frac{\Delta_W}{2 z_m}$ and may be called the ``renormalized'' double-scaling parameter that characterizes the strength of the shock wave.
Note also that $\tau_0=b \sinh^{-1}(t_0/b)$ is the proper time when the left particle emits the shock wave or the $q_W$ quanta.

\subsection{Fast Scrambling and the Comparison with Shock Wave Calculations}

When $t_0$ is negative, we interpret (\ref{4OTOC}) as the effect of the operator $\hat W$ at time $t_0$ on the operator $\hat F$ at time $0$.
When $t_0$ is positive, we interpret it backward in time as the cause at $t_0$ and the effect at time $0$. If we define $\tau_*=b\ln b\bar{\gamma}$, we have
  \be
    \label{F4}
  \langle\hat W(t_0)\hat F(0)\hat W(t_0)\hat F(0)\rangle_{\beta}\propto \left(1+e^{(\tau_*-\tau_0)/b}\right)^{-2\Delta_F} \,.
  \ee
We see that the four-point OTOC is very small as $t_0\rightarrow-\infty$ and doesn't change too much until $\tau_0\simeq\tau_*$.
Thus, $\tau_*$ is identified with the scramble time, after which quantum information gets spread out.
The important property of this system is its high speed of information scrambling.
As $\tau_0$ is positive and significantly greater than $\tau_*$, we find the initial exponential decay of the four-point OTOC when $\tau_0$ passes the thermalization time $b$,
  \be
  \langle\hat W(t_0)\hat F(0)\hat W(t_0)\hat F(0)\rangle_{\beta}\propto 1-2\Delta_Fe^{\tau_*/b}e^{-\tau_0/b}
  \ee
with Lyapunov exponent $\lambda_L=\frac1b=\frac{2\pi}{\beta}$.
These behaviors are illustrated in figure \ref{otoc}.

 These results are consistent with those in \cite{Yeh_23}, but here we do not take the double-scaling parameter $\gamma$ to be small.
As noted in \cite{Murata_17}, when $\bar{\gamma}$ is negative, the four-point OTOC diverges when $\tau_0\simeq\tau_*$, because in this case the negative energy of the shock wave renders the worldsheet wormhole traversable and there emerges a light-like geodesic that connects the two boundaries.
In the picture of eikonal scattering, this corresponds to the non-conservation of probability when $q_W$ quanta have negative energy.

We can also study the late-time behavior of the cross correlator $G_{LR}(t,0)$ for $t\gg b$, by just letting $\tau_2=\tau+ib\epsilon$ in the expression (\ref{overlap3}).
We find that, as $\tau\gg b$,
 \be
 G_{LR}(\tau,0)\propto e^{-\Delta_F \tau/b} \,.
 \ee
Note that $\Delta_F=2$ and $\tau=b\sinh^{-1}(t/b)$. Thus, at late times, we have $G_{LR}(t,0)\propto t^{-2}$, which also agrees with the result obtained by the shock wave method \cite{Yeh_23}.
As noted in \cite{Yeh_23}, when $t\gg b$, the shock wave effect becomes negligible (for the cross correlator without a shock wave, see (\ref{cross}) for the operator $\hat F$).  This can be compared with the four-point OTOC in (\ref{F4}) which indicates that the shock wave effect is important after the scrambling time $\tau_*=b\ln b\bar{\gamma}$ which is supposed to be much longer than the thermalization time $b$.  As can be seen in (\ref{glra}), $G_{LR}(t,0)$ measures the effect of $\hat F$ acting on one boundary at time zero on the same operator at the other boundary at time $t$ in the fixed shock wave background (thus the time of insertion $\hat W(t_0)$ is fixed, whereas in the four-point OTOC we fix the $\hat F$ operators at time zero and vary $t_0$). So, unlike the OTOCs, the two-point function $G_{LR}(t,0)$ is a coarser probe of the shock wave and decays more quickly. It will also be interesting to compare the behavior of the thermal OTOC for $t\ll b$.
However, in this limit, the expression in (\ref{overlap3}) is very sensitive to the choice of the cutoffs $\epsilon_i$. This makes the comparison ambiguous.

\subsection{Six-point OTOCs}

Following \cite{Haehl_21}, we can also apply the eikonal approximation to study the six-point OTOC of the form $\langle\hat W\hat F\hat W\hat W\hat F\hat W\rangle_{\beta}$ for the holographic EPR pair, which encodes the information about scattering behind the worldsheet horizons.
As with the argument of four-point OTOCs in equation (\ref{glra}), this thermal six-point OTOC can be related to the analytical continuation of the boundary correlators
  \be
  \langle TFD|\hat W_L^{\dagger}(t_1)\hat W^{\dagger}_R(t_2)\hat F_L(t_3)\hat F_R(t_4)\hat W_R(t_2)\hat W_L(t_1)|TFD\rangle\equiv F_6 \,.
  \ee
We can write
  \be
   \label{6OTOC}
  F_6=\langle \Psi |\hat F_L(t_3)\hat F_R(t_4)|\Psi \rangle \,,
  \ee
where
  \be
  |\Psi\rangle=\int\int dp_1^Vdp_2^U\Phi_1(p_1^V)\Phi_2(p_2^U)|p_1^Vp_2^U\rangle \,,
  \ee
and the wave functions $\Phi_1$ and $\Phi_2$ for the $q_W$ quanta are the same as the one in (\ref{phi1}) with suitable modifications; taking complex conjugate or replacing $t$ to $-t$, depending on whether there is a hermitian conjugate and whether it is on the left or right boundary.
Here, $|p^Vp^U\rangle$ represents the tensor product of momentum eigenstates of two $q_W$ quanta.
According to the eikonal approximation in this case, we can interpret (\ref{6OTOC}) as two independent scattering events each with their phase factors (see \cite{Maldacena_17} for the details),
   \bea
   &&\langle\tilde{p}_1^V\tilde{p}_2^U|\hat F_L(t_3)\hat F_R(t_4)|p_1^Vp_2^U\rangle\\
   &&\simeq p_1^V\delta(\tilde{p}_1^V- p_1^V)p_2^U\delta(\tilde{p}_2^U- p_2^U)\int\int dq_3^Udq_4^V\langle \hat F_L(t_3)\hat F_R(t_4)|q_3^Uq_4^V\rangle e^{i\delta(s_3)+i\delta(s_4)}
   \eea
where the amplitudes $\delta(s_3)=4\gamma'p_1^Vq_3^U$ and $\delta(s_4)=4\gamma'p_2^Uq_4^V$ are as the ones in the identification we found in (\ref{amplitude}).
$\langle \hat F_L(t_3)\hat F_R(t_4)|q_3^Uq_4^V\rangle=\Phi^F_3(q^U_3)\Phi^F_4(q^V_4)$ is the product of the momentum space wavefunctions for $q_F$ quanta as those in (\ref{phi2}) and (\ref{phi4}).
We then have
 \be
 F_6=\int\int dp^Vdp^U p^Vp^U|\Phi_1(p^V)|^2|\Phi_2(p^U)|^2\int\int dq^U dq^V\Phi^F_3(q^U)\Phi^F_4(q^V) e^{4i\gamma'(p^Vq^U+p^Uq^V)} \,.
 \ee
Note that the $q^U$ and $q^V$ integrations are just the inverse Fourier transforms of bulk-to-boundary correlators back to the position space (see equation (\ref{phi2f})).
After putting all the ingredients together, defining the complexified (proper) time, $\tau_1=\tau_0-ib\epsilon_1$, $\tau_2=\tau_0-ib\epsilon_2$, $\tau_3=ib\epsilon_3$, $\tau_4=ib\epsilon_4$, and redefining the integral contour $p=-4\sin\epsilon_1e^{\tau_0/b}p^V$ and $q=4\sin\epsilon_2e^{\tau_0/b}p^U$, we find
  \bea
  \label{F6int}
  &&F_6=\frac{\pi^6|C_W|^4C_F^2}{\Gamma^4(\Delta_W)2^{2\Delta_F}}
\bigg(\frac{-1}{16 \sin \epsilon_1 \sin \epsilon_2 } \bigg)^{2\Delta_W}
\int_0^{\infty}\int_0^{\infty}dpdq p^{2\Delta_W-1} q^{2\Delta_W-1}\\
  &&\times e^{-p-q}\left(\frac{1}{1+ \gamma'\frac{pe^{i\epsilon_3}}{2\sin\epsilon_1}e^{-\tau_0/b}} \right)^{\Delta_F}\left(\frac{1}{1+\gamma'\frac{qe^{i\epsilon_4}}{2\sin\epsilon_2}e^{-\tau_0/b}}\right)^{\Delta_F} \,.
  \eea
We also assume $\Delta_W\gg\Delta_F=2$. Then $q$ and $p$ integrations are dominated by the regions $p\simeq 2\Delta_W$ and $q\simeq 2\Delta_W$.
To obtain the OTOC of the type $\langle\hat W(t_0)\hat F(0)\hat W(t_0)\hat W(t_0)\hat F(0)\hat W(t_0)\rangle_{\beta}$, we choose again the natural cutoffs $\epsilon_1=\epsilon_2=\frac{z_m}{b}\ll 1$ and set $\epsilon_3=\epsilon_4=0$. We then obtain $F_6$ as an OTOC,
 \be
 \label{F6}
 F_6\propto \left(\frac{1}{1+2 b\bar{\gamma}e^{-\tau_0/b}}\right)^{2\Delta_F}
\,.
 \ee

 \begin{figure}[h]
\centering
\includegraphics[scale=1]{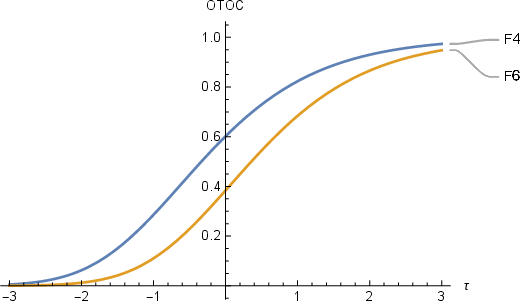}
\caption{We plot the four-point OTOC ($F_4$) and six-point OTOC ($F_6$) in (\ref{F4}) and $(\ref{F6})$ respectively. We rescale the saturated value to be 1, and set $b=1$ , $\tau^*=-2$, (This choice of parameters is just for illustration. In reality, we have$|\tau^*|\gg b$) and $\Delta_F=2$. }
\label{otoc}
\end{figure}

In this case, we define the scrambling time as $\tau'_*=b\ln b2\bar\gamma$. And the six-point OTOC changes significantly only when $\tau_0$ is close to $\tau'_*$. For $\tau_0$ much larger than $\tau'_*$, we see that the six-point OTOC shows an exponential decay with the same Lyapunov exponent, $\lambda_L$, as that of the four-point OTOCs (but the proportional constants are different due to the different scrambling times),
  \be
  F_6\propto 1-2 \Delta_Fe^{\tau'_*/b}e^{-\tau_0/b} \,.
  \ee
Note that by comparing the scrambling time of the six-point OTOC, $\tau'_*$ to the one for four-point OTOC, $\tau_*$, we see that effectively the value of $\bar{\gamma}$ is doubled.
This agrees with the result of \cite{Yeh_23} in the setting of two symmetric shock waves, where the cross correlator is related to the one in one shock wave setting by doubling the value $\gamma$.
It can also be noted that the six-point OTOC has longer scrambling time than the one in four-point OTOCs\footnote{We can see that $\tau'_*-\tau_*=b\ln 2$, which is of the order of thermalization time. This difference vanishes in the infinite-temperature limit, since both the scrambling times approach zero in this limit.} and this is also consistent with the result in \cite{Haehl_21}.
This is because six-point OTOCs probe the more fine-grained chaos \cite{Haehl_18}.
In figure \ref{otoc}, we can see these differences between six-point and four-point OTOCs. Notice that when $\gamma$ is negative, there is also a divergence happens near the scrambling time in the six-point OTOCs. Presumably this divergence has the same origin as that in the four-point OTOCs. 

To study the late-time behavior of $\langle TFD|\hat W_L^{\dagger}(t_0)\hat W^{\dagger}_R(t_0)\hat F_L(t)\hat F_R(0)\hat W_R(t_0)\hat W_L(t_0)|TFD\rangle$ for $t\gg b$, we replace $\epsilon_3$ with $\epsilon_3-i\tau/b$ and take $\tau\gg b$.
The result is
  \be
  F_6\propto e^{-\Delta_F\tau/b}=\left(\frac{t}{b}\right)^{-\Delta_F} \,,
  \ee
which is essentially the cross correlator in the absence of a shock wave. This also agrees with the result in \cite{Yeh_23},
where the effect of shock waves is argued to be negligible for $t \gg b$. The reason that the cross correlator in the two-shock-waves background decays much more quickly than the six-point OTOC is similar to that discussed in the case of four-point OTOCs; namely, OTOCs are finer probes of the shock wave effect compared to two-point correlators.

\section{Discussion}

In this paper, we calculate the four-point and six-point OTOCs for the holographic EPR pair by the worldsheet scattering amplitude in eikonal approximation. We also point out the relation of the worldsheet scattering amplitude with the double scaling parameter, $\gamma$ in the worldsheet shock wave background. By this identification, we argue that there is an equivalence between the scattering amplitude in eikonal approximation and the two-point correlator in the shock wave background even in the worldsheet theory for holographic EPR pairs. We solidify the argument by comparing the four-point and six-point OTOCs obtained here with those obtained by the shock wave method in \cite{Yeh_23}. In the parameter regions when both methods give analytical forms of OTOCs, we found complete agreements.

We observe that the exponential decay rate of the two-point correlator is characterized by the thermalization time scale $b=\beta/2\pi$. The scrambling time of the four-point OTOCs is $b\ln b\bar{\gamma}$, which was estimated as $b\ln S$ in \cite{Yeh_23} with $S$ being the entropy of the wormhole.
Thus, the scrambling time is typically much larger than the thermalization time.
However, unlike the usual duality between an eternal $AdS$ black hole and a one-sided CFT in a thermalized state, where the wormhole entropy is exactly the Von Neumann entropy of the CFT state, here, we interpret the two boundaries of the wormhole as two entangled particles, both of which are physical.
It will then be interesting to know whether the wormhole entropy can be interpreted as the entanglement entropy of the EPR pair. In certain systems, entanglement entropy serves as a more sensitive probe of information spreading compared to OTOCs (e.g., \cite{Harrow_21}). The time dependence of entanglement entropy is influenced by field-field correlators and, in turn, by OTOCs under appropriate conditions (as discussed in the next paragraph). Investigating the relationship between these quantities in the context of the AdS/CFT correspondence could yield further insights.

Furthermore, the EPR pair couples to the ambient CFT, which in the gravitational dual is described by the $AdS_{d+1}$ spacetime, not the string worldsheet.
As studied in \cite{Lee_19}, since the EPR pair does not form a closed system (unlike a one-sided CFT), the particle in the pair actually loses coherence and get disentangled gradually. It will also be interesting to know whether we can see this decoherence in the OTOCs. For this purpose, in the shock wave calculation, we probably need to go beyond the double scaling limit (where the effect of backreaction is completely determined by the coordinate shift in Equation \ref{shift}) and consider the case when two sides of the wormhole have different temperature. In the bulk scattering calculation, this corresponds to going beyond the eikonal approximation.

Another interesting question to ask is how the information insides the horizons is encoded in the EPR pair. One hint may come from the case where the double scaling parameter $\gamma$ is negative, and the worldsheet wormhole becomes traversable.
In this case, the boundary six-point OTOCs may signal the information exchange between the entangled particles through the interior of the black holes. As proposed in \cite{Yeh_23}, this phenomenon could stem from an induced nonlocal interaction. Examining the behavior of six-point OTOCs for $\gamma<0$ might provide insight into the specific type of deformation required to account for this interaction.

\acknowledgments

CPY would like to thank Ben Freivogel for discussion and useful comments on current work during the workshop, Observables in Quantum Gravity, in Aspen. This work was performed in part at Aspen Center for Physics, which is supported by National Science Foundation grant PHY-2210452. The work was also supported in part by National Science and Technology Council (NSTC) of Taiwan under grant number 113-2112-M-259-009 (CPY)
and 111-2112-M-259-006-MY3 (DSL).

\end{document}